\documentclass[reprint,aps,prd,nofootinbib]{revtex4-2}
\usepackage{graphicx}
\usepackage{dcolumn}
\usepackage{bm,amsmath,amssymb}

\usepackage{orcidlink}
\usepackage{natbib}

\usepackage{xcolor}
\newcommand{\sv}{\langle\sigma v\rangle}
\newcommand{\rsp}{r_{\rm sp}}
\newcommand{\rin}{r_{\rm in}}
\newcommand{\gsp}{\gamma_{\rm sp}}
\newcommand{\mdm}{m_{\rm DM}}
\newcommand{\racc}{r_{\rm acc}}
\newcommand{\AO}{A0620$-$00}
\newcommand{\XTE}{XTE J1118$+$480}

\begin{document}

\preprint{APS/123-QED}

\title{Positron escape from dark matter spikes in black hole X-ray binaries} 

\author{Mpho Kgoadi \orcidlink{https://orcid.org/0000-0002-1431-6765}}
\affiliation{School of Physics and Centre for Astrophysics, University of the Witwatersrand, Johannesburg, Wits 2050, South Africa
}%

\author{Geoff Beck \orcidlink{https://orcid.org/0000-0003-4916-4914} }
\altaffiliation{geoffrey.beck@wits.ac.za}
 
\affiliation{School of Physics and Centre for Astrophysics, University of the Witwatersrand, Johannesburg, Wits 2050, South Africa
}%

\date{\today}

             \begin{abstract}
                 Chan and Lee present a model where the AMS-02 positron excess is explained by dark matter (DM) annihilation in spike halos around nearby black holes \AO{} and \XTE{}. Their model assumes a uniform magnetic and diffusive environment en route to Earth. Thus, it depends on the assumption that one can ignore the small-scale, near-horizon environment in the diffusion-loss equation. We demonstrate that this is not the case. An accurate magnetic field model near the horizon means, without increasing the required cross-section by a factor of roughly 500 to 1400, depending on the transport near the horizon, the positron flux from these sources cannot account for a significant fraction of the AMS-02 excess.
             \end{abstract}

\maketitle

\section{Introduction}

PAMELA first reported that the cosmic-ray positron flux exceeded the expected background at energies on the GeV scale~\cite{Adriani2009}. The Alpha Magnetic Spectrometer (AMS-02) subsequently measured the flux with greater precision~\cite{AMS2013,AMS2019}. Its positron-flux measurements can be described as the sum of a diffuse component produced by cosmic-ray interactions with interstellar gas and an unidentified high-energy source component~\cite{AMS2019}. Several studies have proposed astrophysical origins, such as pulsars, for this source~\cite{Hooper2009,CholisKrommydas2022,CholisHoover2023}, whereas others have examined dark matter annihilation as a potential explanation. However, annihilation at the rate needed is constrained by gamma-ray observations of dwarf galaxies~\cite{Hoof2020} as well as by radio observations~\cite{Vollmann2020,Regis2021,Semane2026}.

Chan and Lee~\cite{ChanLee2025} recently suggested a new origin for the unknown source term: dark matter annihilation in density spikes surrounding the two nearest black hole low-mass X-ray binaries, \AO{} and \XTE{}. They report that this explanation fits the AMS-02 spectrum better than the best pulsar models~\cite{ChanLee2025}. A steep density spike develops as a black hole grows at the center of a dark matter distribution~\cite{GondoloSilk1999,Ullio2001,Merritt2002}. Because the annihilation rate is proportional to the density squared, such spikes can turn black holes into bright sources of annihilation products. This possibility has been investigated mainly in the context of supermassive black holes~\cite{GondoloSilk1999,Bertone2002,Fields2014,Lacroix2015,ShapiroShelton2016}. Chan and Lee~\cite{ChanLee2023} extended the idea to \AO{} and \XTE{}, whose orbital periods decay about two orders of magnitude faster than gravitational radiation predicts~\cite{GH2014,GH2017,ChanLee2023}. Magnetic braking~\cite{Justham2006} and circumbinary disks~\cite{ChenLi2015} have both been suggested as explanations, but neither fully agrees with the observations~\cite{ChanLee2023}. Chan and Lee demonstrated that dynamical friction~\cite{Chandrasekhar1943} between the companion star and a dark matter spike can explain the decay. Because the orbits are measured precisely, fitting the data determines the spike index to within a few per cent. Kar et al.~\cite{Kar2024} subsequently used the radio emission of \XTE{} to constrain the annihilation cross section.

With these spikes, Chan and Lee~\cite{ChanLee2025} reproduce the high-energy source term for dark matter with mass $\mdm \approx 8000$~GeV annihilating through the $W^+W^-$ channel. Their fit to the AMS-02 spectrum has a reduced $\chi^2$ of 0.64, compared with $\approx 1.0$ for the best pulsar models~\cite{CholisKrommydas2022}. Their calculation adopts a uniform magnetic field of 2~$\mu$G, corresponding to the value near the Solar System~\cite{SunReich2012}, and assumes this field throughout positron diffusion from the binaries to Earth~\cite{ChanLee2025}. However, that assumption breaks down at the production site. Because the annihilation rate scales as $\rho^2$, most annihilations take place within a few hundred Schwarzschild radii of the black hole, inside a steep spike. The quiescent accretion flow in this region has a magnetic field of order $10^5$~G~\cite{Kimura2021,Kar2024}, eleven orders of magnitude greater than the assumed 2~$\mu$G, and synchrotron losses scale as $B^2$. In this paper, we reproduce Chan and Lee’s result and show that most of the reproduced signal comes from the uniform ambient dark matter density rather than from the spikes. We then replace the uniform field with models for the field near the horizon, and calculate the fraction of the positron signal that escapes the spike. We examine how weak the field would need to be for the proposal to remain viable, and show that the annihilation plateau suppresses the signal regardless of the field strength. Lastly, using the same field, we derive limits on the annihilation cross section from the measured quiescent radio flux of \XTE{} and compare our results with Ref.~\cite{Kar2024}. The paper is organised as follows. Section~\ref{sec:spikes} describes the dark matter spikes; Sec.~\ref{sec:fields} covers the near-horizon magnetic and radiation fields; and Sec.~\ref{sec:propagation} discusses positron transport and emission. Our results are presented in Section~\ref{sec:results}, followed by our conclusions in Sec.~\ref{sec:conclusions}.

\section{Dark matter spike halos} 
\label{sec:spikes}

A dark matter spike forms when a black hole grows adiabatically and compresses the surrounding dark matter. This spike has a density profile $\rho \propto r^{-\gsp}$, where $\gsp \approx 2.25$--$2.5$, according to Ref.~\cite{GondoloSilk1999,Fields2014}, while heating of the dark matter by stellar scattering softens the power law towards $\gsp = 1.5$~\cite{Merritt2004,GnedinPrimack2004}. According to Ref.~\cite{Sadeghian2013}, the density drops to zero at $2 R_s$ instead of $4 R_s$ for a fully relativistic calculation.

Taking the profile form suggested by Chan and Lee using Ref.~\cite{Lacroix2018}, the following function for the dark matter profile is adopted,
\begin{equation}
    \rho_{\text{DM}}(r) =
\left\{
\begin{array}{ll}
    0 &, r \le 2 R_s,\\
    \rho_{0} (r / \rsp)^{-\gsp} &, 2 R_s < r \le \rsp,\\
    \rho_{0} &, r > \rsp,
\end{array}
\right.
\label{eq:profile}
\end{equation}
where $\rho_{0}$ is the ambient galactic density. Following Chan and Lee~\cite{ChanLee2023}, $\rho_0$ is chosen from the Navarro-Frenk-White profile and the final values of $\rho_0$ turn out to be $0.29$ and $0.34$~GeV cm$^{-3}$ for \AO{} and \XTE{}.

Chan and Lee~\cite{ChanLee2023} define the spike radius as $\rsp = 0.2 \rin$, where the radius of influence $\rin$ is given by
\begin{equation}
    M_{\text{DM}}(r \le \rin) = \int_{0}^{\rin} 4 \pi r^2 \rho_{\text{DM}} dr = 2 M_{\text{BH}}.
\label{eq:rin}
\end{equation}

Chan and Lee find $\rin = 5.41$~pc for \AO{} and $5.34$~pc for \XTE{}, giving $\rsp = 1.08$ and $1.07$~pc. The spike index $\gsp$ is derived based on the energy loss due to dynamical friction from the measured period derivative. According to the work in~\cite{ChanLee2023}, the value of $\gsp$ is found to be $1.71^{+0.01}_{-0.02}$ and $1.85 \pm 0.04$ for the \AO{} and \XTE{} cases respectively.

Annihilation injects electrons and positrons at the rate~\cite{ChanLee2025,Vollmann2021}
\begin{equation}
    Q(E, r) = \frac{\sv \rho_{\text{DM}}^2 (r)}{2 \mdm^2} \frac{dN_e}{dE},
\label{eq:source}
\end{equation}
where $\frac{dN_e}{dE}$ refers to the spectrum per annihilation of particles and is obtained from the library PPPC4DMID~\cite{Cirelli2011}. Following Chan and Lee~\cite{ChanLee2025}, the main annihilation channel chosen is $W^+ W^-$ with $\mdm = 8$~TeV. The channel chosen by Kar et al.~\cite{Kar2024}, namely $b \bar{b}$, is used for the comparison in Section~\ref{sec:kar}.

Annihilation also depletes the spike. Over a timescale $t_{\rm BH}$ the density cannot exceed~\cite{Lacroix2018,Vasiliev2007}
\begin{equation}
    \rho_{\text{sat}} = \frac{\mdm}{\sv t_{\text{BH}}}.
\label{eq:rhosat}
\end{equation}

Inside the radius $r_{\text{sat}} = \rsp ( \rho_0 / \rho_{\text{sat}} )^{1 / \gsp}$, where the spike reaches $\rho_{\text{sat}}$, the density no longer follows the steep spike. For an isotropic velocity distribution it rises only as a weak cusp~\cite{Vasiliev2007,ShapiroShelton2016},
\begin{equation}
    \rho_{\text{DM}}(r) = \rho_{\text{sat}} \left( \frac{r_{\text{sat}}}{r} \right)^{1/2},
    \quad \text{for } 2 R_s < r < r_{\text{sat}}.
\label{eq:weakcusp}
\end{equation}

According to~\cite{Lacroix2018,Vasiliev2007}, if the saturated value of the dark matter density is lower than $\rho_0$, then it is possible for the dark matter spike to disappear altogether. Kar et al.~\cite{Kar2024} use the same $\rho_{\rm sat}$, but cap the spike density $\rho_{\rm sp}$ smoothly,
$\rho = \rho_{\rm sp} \rho_{\rm sat} / (\rho_{\rm sp} + \rho_{\rm sat})$,
which produces a flat core.

It is unknown what age the spike has. Based on Ref.~\cite{ChanLee2023,Kar2024}, we can estimate it with the orbital decay time, $t_{\rm BH} \le P/|\dot{P}|$. If the decay of the binary system is the result of dynamical friction on the spike, which is the assumption on which the spike exists, the binary cannot be much older than this. Then for \AO{} we have $4.65\times10^7$~yr and for \XTE{} we obtain $7.73\times10^6$~yr.

\section{Magnetic field models}
\label{sec:fields}

In addition to the field of $B = 2\,\mu$G (that of the vicinity of our Solar system~\cite{SunReich2012}), Chan and Lee have considered a total radiation field made by three blackbody components: CMB, optical and infrared. In total they have obtained $U_{\rm rad} = 1.46$~eV\,cm$^{-3}$. The authors also assumed that the magnetic and radiation fields do not change on propagation from a binary system to Earth~\cite{ChanLee2025}.

Quiescent black hole binaries still accrete at the rate below the Eddington one. Kimura and collaborators~\cite{Kimura2021} model the inner accretion flow as a magnetically arrested disk and use the corresponding model to fit the optical and X-ray spectra of \AO{}, \XTE{} and V404~Cyg. In their approach the field has the form
\begin{equation}
B \simeq 6.2\times10^{5}\,\mathcal{R}_1^{-5/4}\,\dot{m}_{-4}^{1/2}\,M_1^{-1/2}\,
\alpha_{-0.5}^{-1/2}\,\beta_{-1}^{-1/2}~{\rm G},
\label{eq:kimura}
\end{equation}
with $\mathcal{R} = r/R_g$, $R_g = GM_{\rm BH}/c^2$, $\dot{m} = \dot{M}c^2/L_{\rm Edd}$, $\alpha$ the viscosity coefficient, and $\beta$ the plasma $\beta$. Here subscripts denote scaling factors:
$\mathcal{R}_1 = \mathcal{R}/10$, $\dot{m}_{-4} = \dot{m}/10^{-4}$, $M_1 = M_{\rm BH}/(10\,M_\odot)$, $\alpha_{-0.5} = \alpha/10^{-0.5}$, and $\beta_{-1} = \beta/0.1$.
In our simulations we took $\alpha = 0.3$, and the corresponding values from the fits provided in~\cite{Kimura2021}, namely, $\dot{m} = 1.0\times10^{-4}$, $\beta = 0.40$ for the \AO{}, and $\dot{m} = 1.0\times10^{-5}$, $\beta = 0.10$ for the \XTE{}, with the masses of the black holes equal to $M_{\rm BH} = 5.86\,M_\odot$~\cite{vanGrunsven2017}, and $M_{\rm BH} = 7.46\,M_\odot$~\cite{GH2014}.
The slope $\propto r^{-5/4}$ comes from the scalings of the hot flow density $\propto r^{-3/2}$ and the sound speed $\propto r^{-1/2}$ with $r$ at fixed $\beta$.

In a similar way as Kar and coworkers~\cite{Kar2024} we used the model from~\eqref{eq:kimura} for $r < \racc$, and assumed that the field decreases as $(r/\racc)^{-2}$ outside $\racc$, i.e., $r > \racc$. In this approach we did not allow the field to go below the interstellar field of Chan and Lee, which it did at a distance $5.0\times10^{-5}$~pc for \AO{}, and at $3.8\times10^{-5}$~pc for \XTE{}.

Kar and collaborators~\cite{Kar2024} instead take the field at equipartition with the accretion flow. For \XTE{} they adopt $\dot{m} \simeq 9.92\times10^{-5}$, for which their model gives $B_H^{\rm eq} \simeq 4.5\times10^{6}$~G at the horizon.
Scaled to the horizon and the same accretion rate, our model from~\eqref{eq:kimura} is in agreement with it within the 22\% error.
The two estimates differ mainly through the adopted accretion rate, a factor of ten that gives a factor of 3.15 in the field. Kar and coworkers also note that the true field may be up to 20 times below equipartition~\cite{Kar2024}; we test weaker fields in Sec.~\ref{sec:threshold}.
Near the black hole the radiation field has two additional sources, the accretion flow and the companion star. For \AO{}~\cite{Gallo2006,Dincer2018} and for \XTE{}~\cite{Plotkin2015} the corresponding quiescent luminosities are assumed to be $7.1\times10^{30}$~erg\,s$^{-1}$ and $3.5\times10^{30}$~erg\,s$^{-1}$ respectively.
From their effective temperatures and radii, we estimate companion luminosities of $9.7\times10^{32}$ and $3.2\times10^{32}$~erg\,s$^{-1}$, respectively.
The energy loss due to synchrotron radiation and inverse Compton scattering is~\cite{ChanLee2025,Siffert2011}
\begin{equation}
b(r,E) = \left[0.0254\,B_{\mu{\rm G}}^2(r) + 1.02\,U_{\rm rad}(r)\right]\times10^{-16}\,
E_{\rm GeV}^2~{\rm GeV\,s^{-1}},
\label{eq:loss}
\end{equation}
with $B$ in $\mu$G and $U_{\rm rad}$ in eV\,cm$^{-3}$. We write $b = b_0(r)E^2$. The inverse Compton
term assumes the Thomson limit; Klein-Nishina suppression~\cite{BlumenthalGould1970} would only
reduce it, and close to the black hole the losses are dominated by synchrotron radiation.

\section{Positron propagation}
\label{sec:propagation}

Chan and Lee describe the positrons using the steady-state diffusion-loss equation~\cite{ChanLee2025}
\begin{equation}
\frac{D(E)}{r^2}\frac{\partial}{\partial r}\!\left(r^2\frac{\partial n_e}{\partial r}\right)
+ \frac{\partial}{\partial E}\big[b(E)\,n_e\big] + Q(E,r) = 0,
\label{eq:diffloss}
\end{equation}
with $n_e = dn_e/dE$, a diffusion coefficient $D(E) = D_0(E/1~{\rm GeV})^{\delta}$ with
$D_0 = 4\times10^{28}$~cm$^2$\,s$^{-1}$~\cite{WuChen2019} and $\delta = 1/3$, very close to the fitted
$\delta = 0.31 \pm 0.04$ from Ref.~\cite{WuChen2019}, and the losses of
Eq.~\eqref{eq:loss} taken with the constant fields of Sec.~\ref{sec:fields}. Chan and Lee solve this using
the Fourier-series Green's function of Vollmann~\cite{Vollmann2021}, with free escape at a halo
radius of $r_h = 20$~kpc~\cite{Ding2021}, so that
\begin{equation}
\begin{aligned}
n_e(E,d) & = \sum_{n=1}^{\infty} \frac{2}{b(E)\,r_h\,d}\sin\!\left(\frac{n\pi d}{r_h}\right) \\
&\times \int_E^{\mdm}\! dE'\, e^{-n^2[\eta(E)-\eta(E')]} \\
&\times \int_0^{r_h}\! dr'\, r' \sin\!\left(\frac{n\pi r'}{r_h}\right) Q(E',r'),
\label{eq:green}
\end{aligned}
\end{equation}
with $d$ denoting distance to the binary, and $\eta(E) = (\pi^2/r_h^2)\int_E^\infty D(E')/b(E')\,dE'$. The flux at Earth is given via $\Phi = (c/4\pi)\,n_e$. We have used their equation for $\eta(E)$ as published~\cite{ChanLee2025}. Since both $r_h$ and $\sv$ scale the flux, they are effectively fitting the cross section in this way. Their background is given by their Eq.~(9), the diffuse term due to secondaries~\cite{AMS2019,Cavasonza2017}, the values of which parameters we have fitted to the plot in Chan and Lee, as they are not published.

Equation~\eqref{eq:green} applies the interstellar losses everywhere, including inside the spike. In order to take into account the environment near the black hole we follow each individual positron all the way from the production location out to the edge of the spike. Its energy evolves along that trajectory as follows
\begin{equation}
\frac{dE}{dr} = -\,b_0(r)\,E^2\,\frac{dt}{dr}, \qquad dt = \frac{r\,dr}{3\,D(r,E)},
\label{eq:lambda}
\end{equation}
where the time interval follows from a model of the diffusive escape time. We integrate Eq.~\eqref{eq:lambda} from the production radius $r_0$ to $\rsp$ for each injection energy $E_0$, with $D$ following the energy as the positron loses it. The function $E(E_0,r_0)$ provides the final energy at leaving the spike. Positrons are not destroyed, only cooled, so the spectrum transforms with the Jacobian $dE_0/dE$. Each shell is then weighted according to its annihilation rate $\rho_{\rm DM}^2 r^2$, which results in the effective injection spectrum, replacing the injection part in Eq.~\eqref{eq:green}.
The diffusion coefficient close to the black hole is not known. In our nominal model we take Bohm diffusion, $D = r_L c/3$ with $r_L$ the Larmor radius, inside $\racc$, and the interstellar $D(E)$ outside it, where the accretion flow and its turbulent field end. Bohm diffusion is the choice of Kar et al.~\cite{Kar2024}, and is applicable to fields with coherence lengths that are comparable to, or greater than, the gyroradius, per reference~\cite{Kar2024,Regis2023}. Kimura et al.~\cite{Kimura2021}, in turn, use the mean free path $\eta \, r_L$ (where $\eta = 5$) in their accretion flows. In any scenario, our value of $D \le c \, r /3 $ (to give $dt \ge dr /c$) ensures the speed of positron escapes cannot exceed that of light, as we do not know what value $D$ should take in this regime. In Sec.~\ref{sec:threshold} we therefore also test a mean free path of $5 \, r_L$, Bohm diffusion at all radii, and an expression for the Kolmogorov model (here, with $\delta = 1/3$) with coherence length equal to $r$, or $D = (c/3)\,r_L^{1/3} r^{2/3}$. In all cases, we compare with the extreme case in which positrons leave directly along a path at the velocity of light.

For the radio emission we need the electrons and positrons inside the spike. In each shell they cool and escape, and their equilibrium spectrum is governed by $\partial_E [b n_e] - n_e /t_{\rm esc} + 2 Q = 0$, the so-called 'one-zone' model incorporating the escape terms described in~\cite{Kimura2021}. Solving that equation provides us with a means for obtaining
\begin{equation}
\begin{aligned}
\frac{dn_e}{dE}(r,E) &= \frac{1}{b(r,E)} \int_{E}^{\mdm}\! dE'\, 2Q(E',r) \\
&\quad \times \exp\!\left[-\int_{E}^{E'} \!\frac{dE''}{b(r,E'')\,t_{\rm esc}(r,E'')}\right].
\end{aligned}
\label{eq:equilibrium}
\end{equation}
where the '2' is due to counting both the electrons and positrons, and in addition $t_{\rm esc} = r^2/(6D)$. Every particle injected above $E$ cools through $E$, and the exponential is the probability that it has not escaped while cooling from $E'$ to $E$. In particular, if $t_{\rm esc} \to \infty$, this again becomes the 'loss only' form of the equation commonly used to compute the spike emission~\cite{Kar2024,Lacroix2015,RegisUllio2008}. Particles that escape a shell are not followed to larger radii, where the magnetic field is smaller.

For the power of a single electron, synchrotron radiation is given as~\cite{RybickiLightman1979}
\begin{equation}
P(\nu,E) = \int_{0}^{\pi} \frac{\sqrt{3}\,e^3 B \sin^2 \alpha}{2 m_e c^2}
    F \!\left(\frac{\nu}{\nu_c \sin \alpha} \right) d\alpha,
\label{eq:psync}
\end{equation}
with $\nu_c = \tfrac{3}{2} \gamma^2 eB / (2\pi m_e c)$ and $F(x) = x \int_{x}^\infty K_{5/3}(t)\, dt$. As in~\cite{DarkMatters}, we again integrate over $\alpha$ explicitly rather than using the already angle-averaged form and, in particular, use $F(x) \simeq 2.1495\,x^{1/3}$ for $x < 10^{-3}$. Flux is then given as 
\begin{equation}
S_\nu = \frac{1}{4\pi d^2} \int_{2 R_s}^{\rsp} 4\pi r^2\, dr \int dE\, \frac{dn_e}{dE} \, P(\nu,E).
\label{eq:snu}
\end{equation}

For each target we compute $\chi^2(\sv) = \sum_i [S_\nu^{\rm model}(\nu_i) - S_i]^2/\sigma_i^2$
over the measured fluxes $S_i \pm \sigma_i$ and exclude, at 95\% confidence, the cross sections with
$\Delta\chi^2 > 2.71$ relative to the minimum, keeping only those that over-predict the data. A model
that falls below the data cannot be excluded, since the jet, which we do not model, can provide the
rest. Because $\sv$ also sets $\rho_{\rm sat}$, the predicted flux is not monotonic in $\sv$, so we scan
$\sv$ and quote the lowest excluded value. For \XTE{} we use the quiescent radio flux density measured
with the Karl G. Jansky Very Large Array, $4.79 \pm 1.45~\mu$Jy at 5.3~GHz~\cite{Gallo2014}, the same
measurement used by Kar et al.~\cite{Kar2024}.

\section{Results}
\label{sec:results}

\subsection{Reproducing the results of Chan and Lee}
\label{repro}

\begin{figure}
\includegraphics[width=\columnwidth]{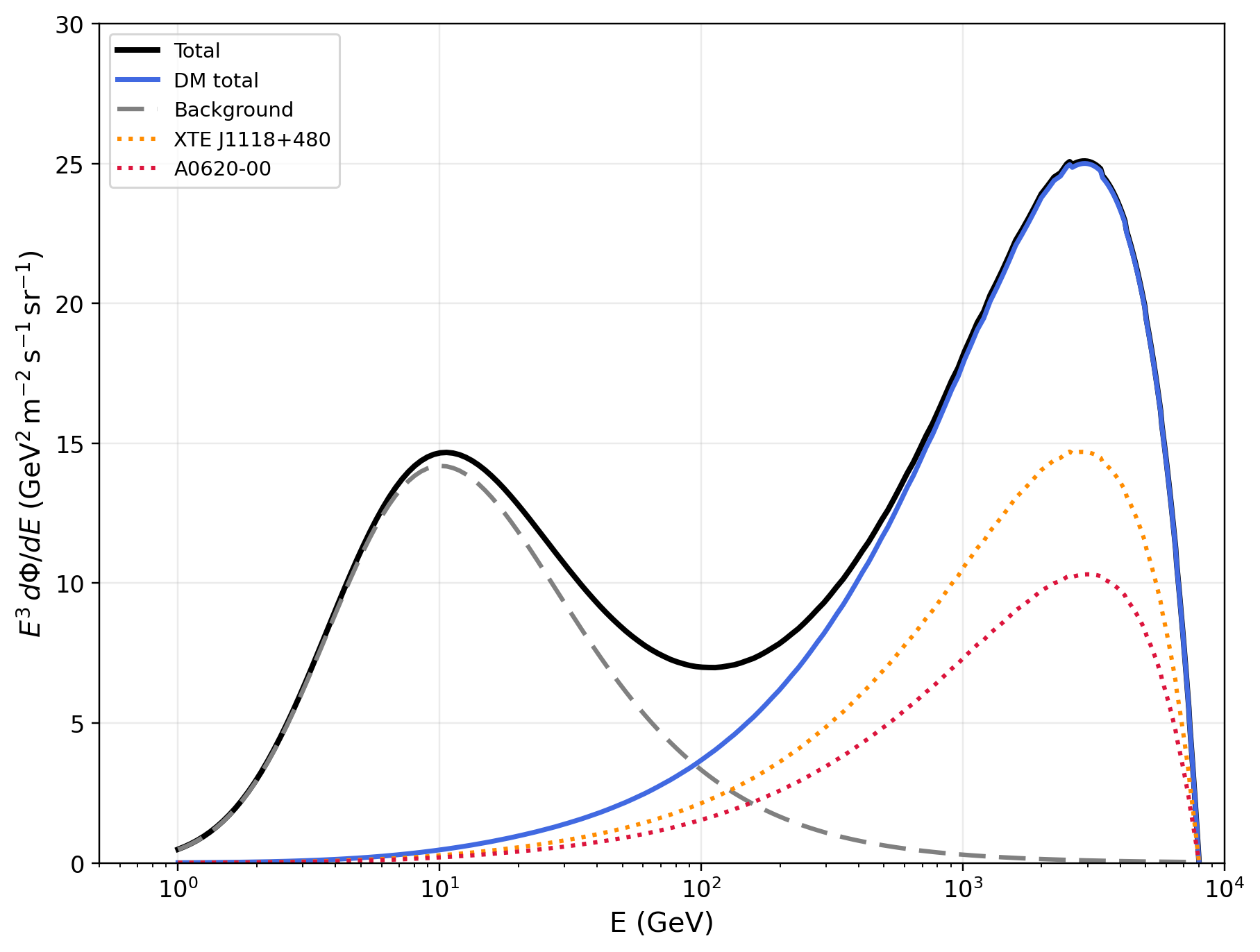}
\caption{Our reproduction of Fig.~1 of Chan and Lee~\cite{ChanLee2025}: the positron flux at Earth,
$E^3\,d\Phi/dE$, for $\mdm = 8$~TeV annihilating into $W^+W^-$, with $B = 2\,\mu$G throughout. Dashed:
the cosmic-ray background; dotted: the contributions of the two binaries; blue: their sum; black:
the total.}
\label{fig:repro}
\end{figure}

We start by reproducing the results from \cite{ChanLee2025} with their assumptions, namely the DM spike profile from \cite{ChanLee2023}, the positron yields from annihilation into $W^+W^-$ from \cite{Cirelli2011} at $\mdm = 8000$~GeV, and their Fourier series solution to the diffusion-loss equation \cite{Vollmann2021} with $B = 2 \mu$G and $U_{\rm rad} = 1.46$~eV\,cm$^{-3}$ everywhere. The background from cosmic rays is from their eq. (9). The results are shown in Fig.~\ref{fig:repro}.

The DM contribution peaks around 3000~GeV and matches what is shown in \cite{ChanLee2025}. To reproduce the high energy excess, we need $\sv = 8.2 \times 10^{-22}$~cm$^3$\,s$^{-1}$, $2.7 \times 10^4$ times the thermal value. They refer to this as the effective annihilation cross section. From Fig.~\ref{fig:repro}, \XTE{} gives a larger contribution than \AO{} despite being further away.

Separating the source into its two parts shows where this signal comes from. Eq.~\eqref{eq:profile} assigns the constant density $\rho_0$ to all radii beyond $\rsp$, and the radial integral in Eq.~\eqref{eq:green} extends to $r_h$, so each binary is surrounded by an ambient region of uniform density that fills the diffusion halo. At the peak this region supplies 97.9\% of the flux and the spikes only 2.1\%. The ambient region alone fits the excess with $\sv = 8.4\times10^{-22}$~cm$^3$\,s$^{-1}$, while the spikes alone require $3.06\times10^{-20}$~cm$^3$\,s$^{-1}$, 37 times more. This is also why \XTE{} contributes more than \AO{}: the ratio of the two contributions, 1.42, is close to the ratio of the squared ambient densities, $(0.34/0.29)^2 = 1.37$, whereas for the spikes alone it would be about 270. The ambient term is not associated with the black holes and would be present without them. The proposal of Ref.~\cite{ChanLee2025} rests on the spikes, and in the rest of this section we consider them alone.

\subsection{Near-horizon environment}
\label{sec:environment}

Chan and Lee make an assumption that the magnetic field $B$ is $2 \mu$G everywhere, including within a few hundred Schwarzschild radii of the horizon, where the positrons are produced. We replace this with the magnetically arrested accretion flow of Kimura et al.~\cite{Kimura2021}, which is fitted to the quiescent spectral distributions of \AO{} and \XTE{}. At $10\,R_g$ the field is
$4.2\times10^5$~G for \AO{} and $2.3\times10^5$~G for \XTE{}. It falls as $r^{-5/4}$ inside the
accretion radius $r_{\rm acc} = 0.6\,a$, with $a$ the orbital separation, and as $r^{-2}$ beyond
it, following Kar et al.~\cite{Kar2024}, reaching the interstellar 2~$\mu$G only at
$5.0\times10^{-5}$~pc and $3.8\times10^{-5}$~pc respectively. Scaled to the horizon and to the
accretion rate of Ref.~\cite{Kar2024}, it agrees with their independent equipartition estimate,
$4.5\times10^6$~G, to 22\%.

\begin{figure*}
\includegraphics[width=\textwidth]{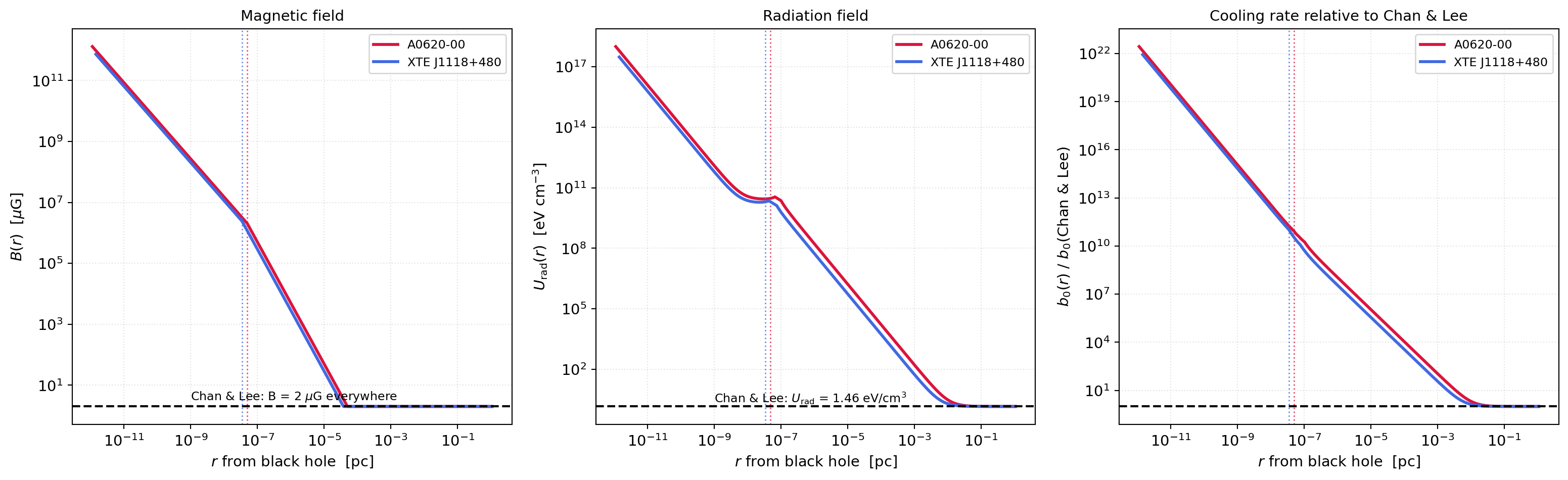}
\caption{The near-horizon environment of \AO{} (red) and \XTE{} (blue). Left: the magnetic field of
Kimura et al.~\cite{Kimura2021} with the radial profile of Kar et al.~\cite{Kar2024}; the dashed line
is the constant 2~$\mu$G of Chan and Lee. Centre: the radiation field. Right: the cooling coefficient
relative to the constant interstellar value used by Chan and Lee. Dotted vertical lines mark the
accretion radius $r_{\rm acc}$.}
\label{fig:environment}
\end{figure*}

Figure~\ref{fig:environment} shows the field, the radiation field, and the resulting cooling
coefficient relative to the one assumed by Chan and Lee. At $10^{-8}$~pc the positrons lose energy
$3.7\times10^{12}$ (\AO{}) and $2.1\times10^{12}$ (\XTE{}) times faster than in their calculation,
and still $1.1\times10^{8}$ and $3.6\times10^{7}$ times faster at $10^{-6}$~pc.

The consequence for escape follows from Eq.~\eqref{eq:lambda}. Even for the fastest possible escape, in
a straight line at the speed of light, the energy on leaving the spike cannot exceed
$1/\Lambda$ with $\Lambda = \int b_0\,dr/c$. A positron produced at $r_0 = 10^{-9}$~pc therefore leaves with
at most 80~GeV (\AO{}) or 140~GeV (\XTE{}), and one produced at $10^{-8}$~pc with at most 2.6~TeV or
4.8~TeV, whatever its injection energy. Any slower transport leaves it with less. With the fields of Chan and Lee, by
contrast, an 8~TeV positron loses only a fraction $3\times10^{-6}$ of its energy inside the spike, so that
cooling there is negligible.

\subsection{Effect on the positron flux}
\label{sec:fieldeffect}

\begin{figure*}
\includegraphics[width=\textwidth]{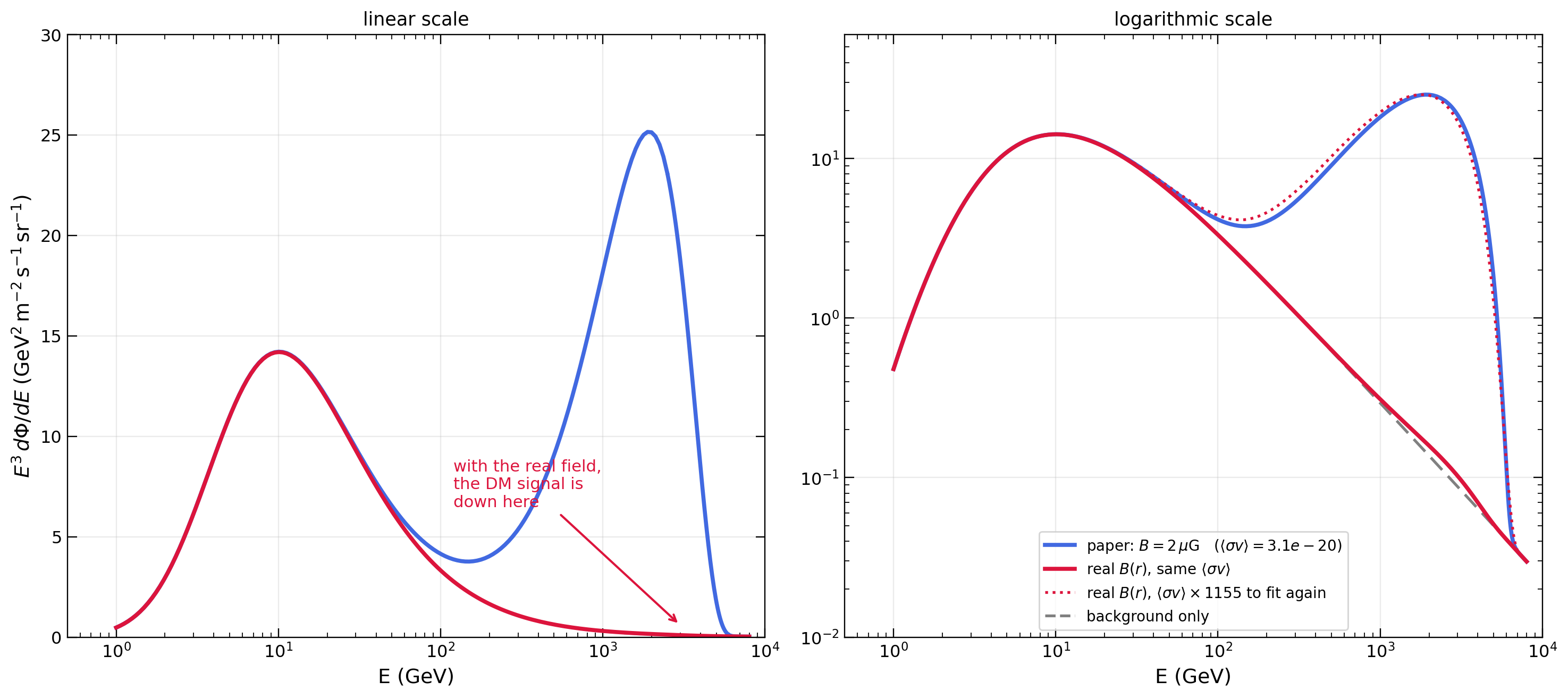}
\caption{The positron flux at Earth for the spike alone, computed twice. Blue: $B = 2\,\mu$G, as
assumed by Chan and Lee. Red: the near-horizon field of Fig.~\ref{fig:environment}, at the same
cross section. Dotted red: the near-horizon field with the cross section raised by a factor of 1155
to recover the peak. Dashed: the background. Left: linear scale, as in Ref.~\cite{ChanLee2025};
right: logarithmic scale.}
\label{fig:fieldeffect}
\end{figure*}

In order to isolate the effect of the field, we repeat the same calculation twice and change only $B$. Following Sec.~\ref{repro}, the source is the spikes alone, $2R_s < r \le \rsp$, without annihilation saturation, which is the case most favourable to the proposal. With $B = 2\,\mu$G, fitting the peak requires $\sv = 3.06\times10^{-20}$~cm$^3$\,s$^{-1}$.
With the near-horizon field, the same cross section gives a peak 1155 times lower
(Fig.~\ref{fig:fieldeffect}), and restoring it requires $\sv = 3.5\times10^{-17}$~cm$^3$\,s$^{-1}$,
about $10^9$ times the thermal value.

The suppression depends only weakly on energy: between 0.13\% (at 10~GeV) and 0.06\% (at 5~TeV) of the
2~$\mu$G flux survives. The field therefore lowers the signal almost without reshaping it.
Positrons produced in the strong-field region are not destroyed; they all escape eventually, but at
energies far below the TeV range of the excess. The size of the suppression is largely set by the
fraction of annihilations that occur outside $\racc$, where the positrons can still escape with their
energy: $1.1\%$ for \AO{} and $0.086\%$ for \XTE{}, which dominates the flux from the spike alone.
Because the factor of 1155 is a ratio of two calculations on the same spike, it does not depend on the
spike normalisation. In the full model of Sec.~\ref{repro}, where the spikes supply 2\% of the peak, the
near-horizon field lowers the peak by only 2\%; with the field included, the spikes supply a fraction
$2\times10^{-5}$ of the flux needed to explain the excess.

The strong-field region is small. The field exceeds 1~kG only within $620\,R_s$ (\AO{}) and
$390\,R_s$ (\XTE{}), a fraction of about $10^{-29}$ of the spike volume, and it is natural to
neglect it. However, the annihilation rate per radial shell scales as
$\rho^2 r^2\,dr \propto r^{2-2\gsp}\,dr$, which for $\gsp > 1.5$ is dominated by the innermost radii.
In the profile of Chan and Lee this small region contains 91\% (\AO{}) and 98\% (\XTE{}) of all
annihilations. The positrons are produced where the field is strongest.

\subsection{How weak can the field be?}
\label{sec:threshold}

To test how robust this conclusion is, we repeat the calculation of Sec.~\ref{sec:fieldeffect}
with the near-horizon field scaled down by a factor $f$, keeping the interstellar floor,
$B_f(r) = \max[f\,B(r),\,2\,\mu{\rm G}]$. We do this for each of the transport models of
Sec.~\ref{sec:propagation}, and we set the radiation field to its interstellar value, which favours
Chan and Lee.

At the nominal field the suppression ranges from 473, for positrons leaving in a straight line at the speed of
light, to 1374, for Bohm diffusion everywhere (Table~\ref{tab:models}). It is 790 for a Kolmogorov
form with $\delta = 1/3$, close to the Galactic value, so the conclusion does not depend on taking Bohm
diffusion. The suppression is reduced only if the positrons move faster than light, as they would if the
interstellar $D(E)$ were used all the way to the horizon: its mean free path, more than a parsec, is far larger
than the region itself, and diffusion does not apply.

\begin{figure}
\includegraphics[width=\columnwidth]{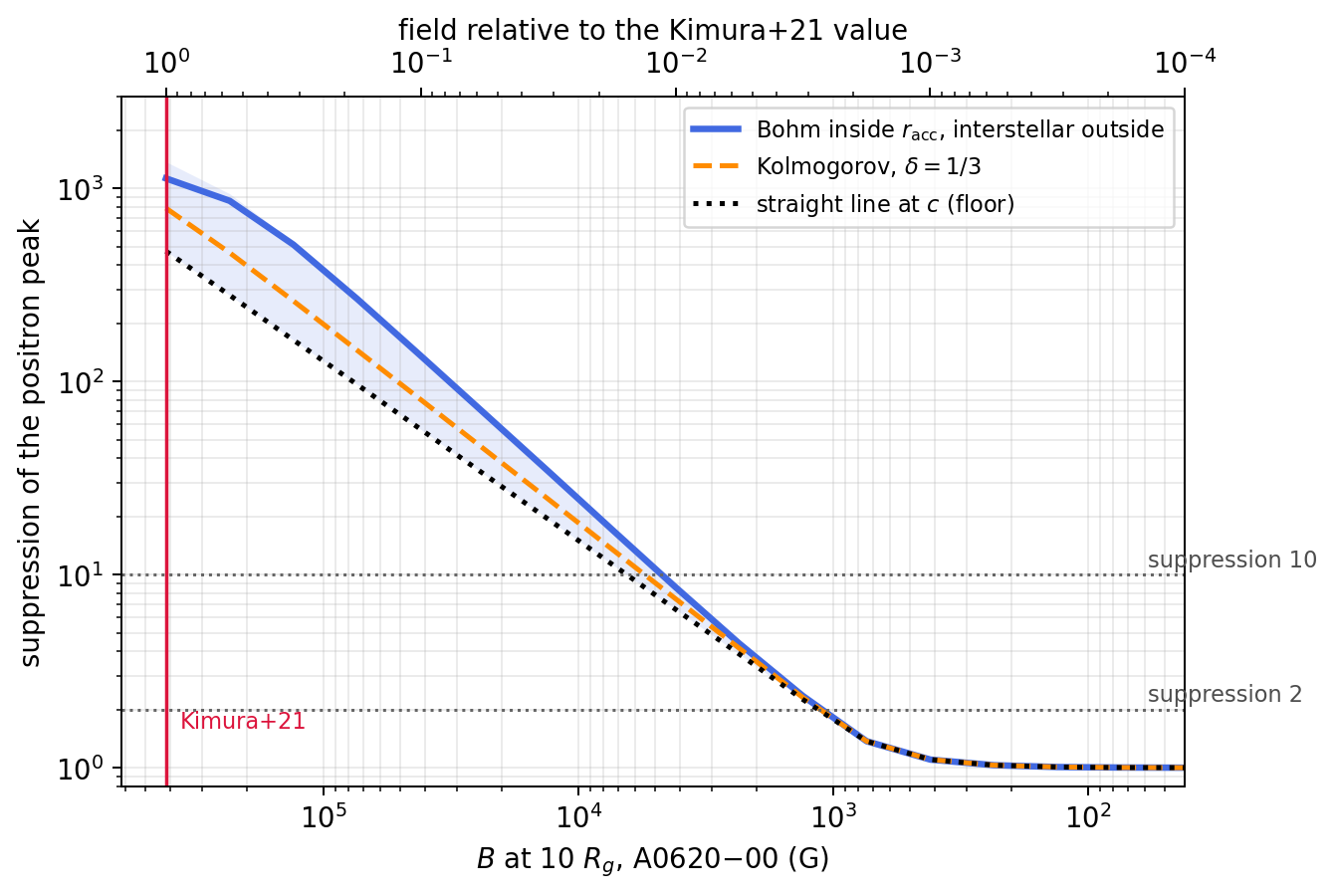}
\caption{Suppression of the positron peak as the near-horizon field is weakened, against the field
at $10\,R_g$ for \AO{} (bottom axis) and relative to the field of Kimura et al.~\cite{Kimura2021}
(top axis, red line). Solid: nominal transport; dashed: Kolmogorov form with $\delta = 1/3$; dotted:
straight-line escape at the speed of light. The band spans all the models of Table~\ref{tab:models}.}
\label{fig:threshold}
\end{figure}

The suppression falls only slowly as the field weakens (Fig.~\ref{fig:threshold} and
Table~\ref{tab:threshold}). In the nominal model it is still a factor of 100 for a field 13 times weaker, a
factor of 10 for a field 88 times weaker, and a factor of 2 only for a field 376 times weaker. Across the
transport models these become 5 to 13, 64 to 89 and 364 to 383. The result of Chan and Lee therefore
survives only if the field at $10\,R_g$ is below about $10^3$~G, while the two independent estimates of
Sec.~\ref{sec:fields} give $\sim 10^5$~G. Kar et al.\ note that observations may allow fields up to 20
times below equipartition~\cite{Kar2024}; such a field still suppresses the peak by a factor of 30 to 60. In
terms of the plasma parameter, $B \propto \beta^{-1/2}$, a suppression of only 2 requires $\beta \gtrsim 10^4$, against $\beta = 0.1$--$0.4$ in the fits of Ref.~\cite{Kimura2021}.

\begin{table*}
\caption{Suppression $S$ of the positron peak for each transport model, at the nominal field and for a
field 20 times weaker, and the factor $1/f$ by which the field must be weakened to reach $S = 100$, 10 and 2.
All models are capped at $D \le c\,r/3$.}
\label{tab:models}
\begin{ruledtabular}
\begin{tabular}{lccccc}
Model & $S(1)$ & $S(1/20)$ & $S=100$ & $S=10$ & $S=2$ \\
\hline
Bohm inside $\racc$ & 1127 & 59 & 13 & 88 & 376 \\
Mean free path $5\,r_L$ & 895 & 33 & 8 & 64 & 364 \\
Bohm everywhere & 1374 & 60 & 13 & 89 & 383 \\
Kolmogorov, $\delta = 1/3$ & 790 & 39 & 8 & 76 & 371 \\
Straight line at $c$ & 473 & 30 & 5 & 64 & 364 \\
\end{tabular}
\end{ruledtabular}
\end{table*}

\begin{table*}
\caption{The near-horizon field needed for the positron peak of Chan and Lee to be suppressed by a
given factor. $f$ is the field relative to that of Eq.~\eqref{eq:kimura}, for the nominal transport model.}
\label{tab:threshold}
\begin{ruledtabular}
\begin{tabular}{cccc}
Suppression & $1/f$ & $B(10\,R_g)$ [G], \AO{} / \XTE{} & $\beta$, \AO{} / \XTE{} \\
\hline
1127 & 1 & $4.2\times10^5$ / $2.3\times10^5$ & 0.4 / 0.1 \\
100 & 13 & $3.2\times10^4$ / $1.8\times10^4$ & 68 / 17 \\
10 & 88 & $4.7\times10^3$ / $2.6\times10^3$ & $3.1\times10^3$ / $7.7\times10^2$ \\
2 & 376 & $1.1\times10^3$ / 620 & $5.7\times10^4$ / $1.4\times10^4$ \\
\end{tabular}
\end{ruledtabular}
\end{table*}

\subsection{The annihilation plateau}
\label{sec:plateau}

A second effect, independent of the magnetic field, acts at the source: the annihilation plateau of
Eqs.~\eqref{eq:rhosat} and~\eqref{eq:weakcusp}, with the spike ages of Sec.~\ref{sec:spikes}.

\begin{figure*}
\includegraphics[width=\textwidth]{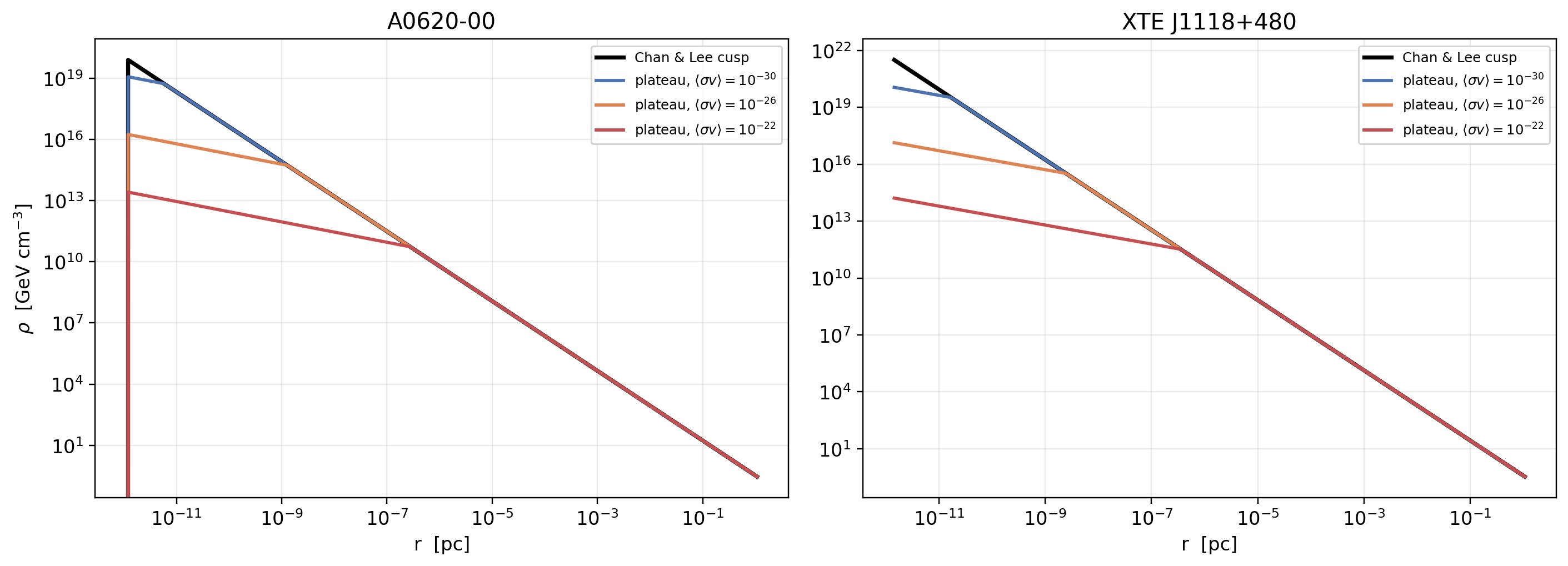}
\caption{The dark matter density around \AO{} (left) and \XTE{} (right). Black: the spike of Chan and
Lee. Coloured: the same spike with the annihilation plateau, for three cross sections, with the spike
age set by the orbital decay time.}
\label{fig:plateau}
\end{figure*}

At $2R_s$ the spike of Chan and Lee reaches $9.0\times10^{19}$~GeV\,cm$^{-3}$ (\AO{}) and
$3.1\times10^{21}$~GeV\,cm$^{-3}$ (\XTE{}), so the plateau is avoided only for
$\sv < 6\times10^{-32}$ and $1\times10^{-32}$~cm$^3$\,s$^{-1}$. Figure~\ref{fig:plateau} shows the
profiles, and Table~\ref{tab:plateau} the resulting suppression of the annihilation rate,
$\mathcal{R} = \sv/(2\mdm^2)\int\rho^2\,dV$. At the effective cross section $\sv = 2\times10^{-22}$~cm$^3$\,s$^{-1}$, the rate
falls by factors of 180 and 5760, and the reduction grows with $\sv$. This involves no magnetic field, propagation or emission model.

\begin{table}
\caption{Suppression of the annihilation rate by the plateau, for $\mdm = 8$~TeV.}
\label{tab:plateau}
\begin{ruledtabular}
\begin{tabular}{lcc}
& $\sv = 3\times10^{-26}$ & $\sv = 2\times10^{-22}$ \\
& cm$^3$\,s$^{-1}$ & cm$^3$\,s$^{-1}$ \\
\hline
\AO{} & 21 & 180 \\
\XTE{} & 206 & 5760 \\
\end{tabular}
\end{ruledtabular}
\end{table}

\subsection{Comparison with Kar et al.}
\label{sec:kar}

\begin{figure}
\includegraphics[width=\columnwidth]{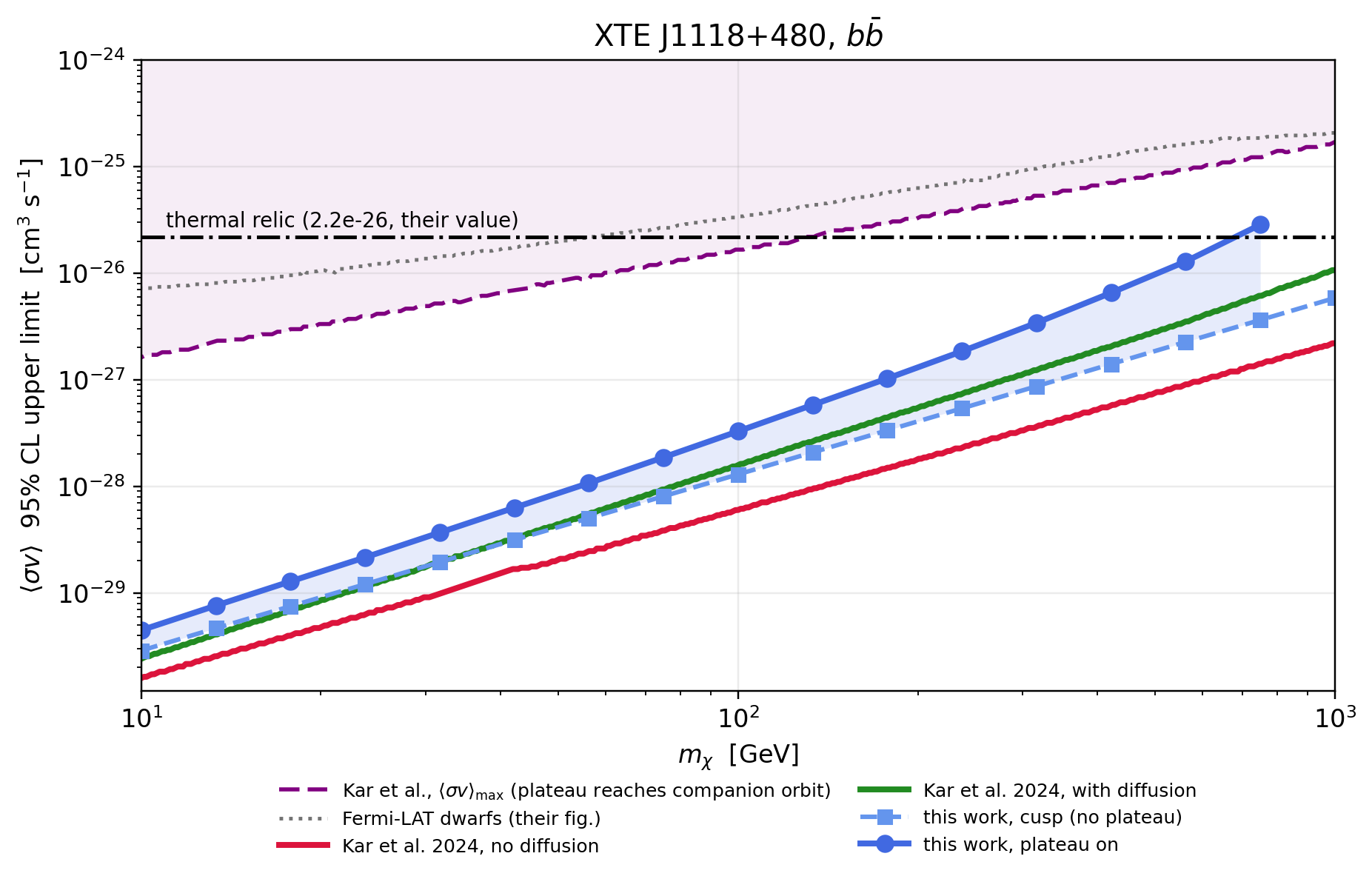}
\caption{95\% upper limits on $\sv$ for \XTE{} in the $b\bar{b}$ channel. Red and green: Kar et
al.~\cite{Kar2024} without and with diffusion, digitised from their Fig.~5. Blue: this work, with
(solid, ending where the low-cross-section branch ends) and without (dashed) the annihilation plateau. Purple: the largest $\sv$ for which the
plateau stays inside the companion's orbit~\cite{Kar2024}. Dotted: Fermi-LAT dwarf
galaxies, from the same figure. Dash-dotted: the thermal cross section.}
\label{fig:kar}
\end{figure}

Kar et al.~\cite{Kar2024} bound $\sv$ from the radio emission of \XTE{} alone, in the $b\bar{b}$
channel, with an independent implementation. We compute the same limit (Fig.~\ref{fig:kar}). They use the
same $\rho_{\rm sat}$ and spike age as we do, but a flat core rather than the weak cusp of
Eq.~\eqref{eq:weakcusp}, so our limit with the plateau is the closest comparison. From 10 to 750~GeV the
plateau weakens the limit by a factor of 1.6, rising to 7.8. Our limit falls below the thermal value,
$2.2\times10^{-26}$~cm$^3$\,s$^{-1}$~\cite{Steigman2012}, for $m_\chi < 680$~GeV (320~GeV for $W^+W^-$).
Above these masses (750~GeV for $b\bar{b}$, 400~GeV for $W^+W^-$), the plateau keeps the predicted flux at
low cross sections below the data, and the first excluded cross section jumps to $\sim 10^{-22}$~cm$^3$\,s$^{-1}$,
so our curve ends there.

Our limit is 1.9 to 4.7 times weaker than their diffusive curve. Most of this comes from two inputs that
neither paper states as a convention (Table~\ref{tab:kar}). Both use $\rsp = 0.2\,r_{\rm in}$, but Kar et
al.\ obtain $r_{\rm in} \simeq 7.3$~pc~\cite{Kar2024} where Chan and Lee obtain 5.34~pc~\cite{ChanLee2023}.
Since the limit scales as $\rsp^{-2\gsp}$ (we find $\rsp^{-4.1}$ numerically), this factor of 1.37 in
$\rsp$ is a factor of $\approx 3.7$ in the limit. Independently, Kar et al.\ take an accretion rate of
$9.92\times10^{-5}$ of the Eddington rate where we take $1.0\times10^{-5}$~\cite{Kimura2021}, a factor of
3.15 in $B$. At 100~GeV the two choices move the limit in opposite directions, while at 500~GeV the
accretion rate has little effect. Matching both places our limit between their two curves, within 15\% of
the diffusive one at 100~GeV and 4\% at 500~GeV.

\begin{table*}
\caption{The two inputs that separate our limit for \XTE{} ($b\bar{b}$) from that of Kar et
al.~\cite{Kar2024}: the limit, in cm$^3$\,s$^{-1}$, and its ratio to their limit with diffusion.}
\label{tab:kar}
\begin{ruledtabular}
\begin{tabular}{lcccc}
& \multicolumn{2}{c}{$m_\chi = 100$ GeV} & \multicolumn{2}{c}{$m_\chi = 500$ GeV} \\
Configuration & limit & ratio & limit & ratio \\
\hline
This work & $3.29\times10^{-28}$ & 2.08 & $9.75\times10^{-27}$ & 3.46 \\
$\rsp = 1.46$ pc & $8.33\times10^{-29}$ & 0.53 & $1.96\times10^{-27}$ & 0.69 \\
$\dot{m} = 9.92\times10^{-5}$ & $4.89\times10^{-28}$ & 3.09 & $1.09\times10^{-26}$ & 3.87 \\
Both & $1.35\times10^{-28}$ & 0.85 & $2.70\times10^{-27}$ & 0.96 \\
\end{tabular}
\end{ruledtabular}
\end{table*}

\section{Conclusions}
\label{sec:conclusions}

Chan and Lee~\cite{ChanLee2025} proposed that dark matter annihilating in density spikes around
\AO{} and \XTE{} produces the high-energy positron excess measured by AMS-02. Their calculation
propagates the positrons through a constant 2~$\mu$G field, including the region a few hundred Schwarzschild
radii from the horizon where they are produced. We have reproduced their result and replaced this
assumption with models of the near-horizon field fitted to these two binaries.

In their model as written, about 98\% of the reproduced positron peak comes from the uniform ambient
density assigned to the diffusion halo around each binary, and only about 2\% from the spikes, which
alone require a cross section 37 times larger.

Although the strong-field region is a tiny fraction of the spike volume, it contains over 90\% of
the annihilations, because the rate scales as $\rho^2$. In that field the positrons lose almost all their
energy before leaving the spike. At the same cross section, the predicted positron peak falls by a factor of
roughly 500 to 1400, depending on the transport near the horizon. The lower value is for positrons leaving in a
straight line at the speed of light, and so does not depend on any assumption about diffusion. The result is
robust: the field would have to be about 400 times weaker than two independent estimates for the proposal to
survive, corresponding to a plasma parameter $\beta \gtrsim 10^4$. Independently of the field, annihilation caps
the density of the spike. With an age set by the orbital decay, this lowers the annihilation rate by two to four
orders of magnitude at the cross section the proposal requires. Either effect alone rules out the proposed
explanation of the AMS-02 excess. The same field makes the binaries synchrotron sources. The measured
quiescent radio flux of \XTE{} constrains $\sv$ below the thermal value up to $m_\chi \approx 680$~GeV for $b\bar{b}$. Matching
the spike radius and accretion rate of Kar et al.~\cite{Kar2024} brings our bound to within 15\% of theirs, and
these two inputs, which neither paper states as a convention, account for most of the difference.

\clearpage

\bibliography{refs}

\end{document}